# THERMODYNAMIC PROPERTIES OF A NUCLEON UNDER THE GENERALIZED SYMMETRIC WOODS-SAXON POTENTIAL IN FLOURINE 17 ISOTOPE

**Bekir Can LÜTFÜOĞLU[1,*], Muzaffer ERDOGAN[2]**

[1] Department of Physics, Faculty of Science, Akdeniz University, Antalya, Turkey
[2] Department of Physics, Faculty of Science and Letters, Namik Kemal University, Tekirdag, Turkey

## ABSTRACT

The exact analytical solution of the Schrödinger equation for a generalized symmetrical Woods-Saxon potential are examined for a nucleon in Fluorine 17 nucleus for bound and quasi-bound states in one dimension. The wave functions imply that the nucleon is completely confined within the nucleus, i.e., no decay probability for bound states, while tunneling probabilities arise for the quasi-bound state. We have calculated the temperature dependent Helmholtz free energies, the internal energies, the entropies and the specific heat capacities of the system. It is shown that, when the quasi-bound state is included, the internal energy and entropy increase, while the Helmholtz energy decreases at high temperatures. Very high excitation temperatures imply that the nucleus does not tend to release a nucleon. The calculated quasi bound state energy is in reasonable agreement with the experimental data on the cumulative fission energy issued by IAEA.

Corresponding Author: bclutfuoglu@akdeniz.edu.tr

## 1. INTRODUCTION

Recently, the interest to the thermodynamic functions, which play significant role to understand fully the physical properties of different potential fields in either relativistic or non-relativistic regimes, has been increased. Pacheco et al. have analyzed one [1] and three dimensional [2] Dirac oscillator in a thermal bath. Boumali has studied relativistic harmonic oscillator in context of thermodynamics [3], calculated the thermal properties of graphene under a magnetic field via the two dimensional Dirac oscillator [4] and thermodynamic properties of the one-dimensional Duffin-Kemmer-Petiau oscillator by using the Hurwitz zeta function method [5]. Arda et al. have studied the thermodynamic quantities of linear potential with Klein-Gordon and Dirac equations [6]. Dong et al. have studied the thermodynamic properties of a non-relativistic harmonic oscillator with an inverse square potential [7].

On the other hand, the Woods-Saxon (WS) potential has numerous applications in physics such as nuclear physics [8-12], atom-molecule [13] and for non-relativistic [12-16] and relativistic problems [17-25].

While WS potential can be used to model the behavior of the nuclear force, which rapidly decays to zero for large radii, the large force exerted on the nucleon near the surface can be represented by an additional term to WS potential. This model is called the Generalized Symmetric Woods-Saxon (GSWS) potential [11].

GSWS potential can be used to model any system, in which a particle is confined in a potential well enclosed by a potential barrier, such as a nucleon in an atomic nucleus, an electron in a metal, the scattering, transmission resonance, supercriticality, decay, fusion, fission [26-30].

In this paper, we are motivated to calculate the thermodynamic properties of a nucleon in a Fluorine 17 nucleus by using the GSWS potential instead of the WS potential. Fluorine element has a wide usage in nuclear medical monitoring, such as positron emission tomography–computed tomography (PET-CT) [31]. Although Fluorine has 18 isotopes, Fluorine 17 isotope is the only one that can emits positron. Hence, the thermodynamic properties of the isotope worth to investigate.

In Section II, we discussed the characteristics of the GSWS potential and give the energy spectrum of the nucleon obtained by using the method in [14]. In Section III, we first calculate the partition function based on the energy spectrum, then, the temperature dependent Helmholtz free energies, the internal energies, the entropies and the specific heat capacities of the system. Finally in Section IV, our conclusion is given.

## 2. METHODS

To study the behavior of a nucleon confined within one dimensional potential well, we first consider the Woods Saxon potential;

$$V_{WS}(x) = -\theta(-x)\frac{V_0}{1+e^{-a(x+L)}} - \theta(x)\frac{V_0}{1+e^{a(x-L)}}, \quad (1)$$

In order to include the surface effect, we add a second term, the radius times the derivative of the WS potential with respect to position. We obtain GSWS potential as;

$$V(x) = \theta(-x)\left[-\frac{V_0}{1+e^{-a(x+L)}} + \frac{V_0 aLe^{-a(x+L)}}{(1+e^{-a(x+L)})^2}\right] + \theta(x)\left[-\frac{V_0}{1+e^{a(x-L)}} + \frac{V_0 aLe^{a(x-L)}}{(1+e^{a(x-L)})^2}\right], \quad (2)$$

here $\theta(\pm x)$ are the Heaviside step functions. The surface term corresponds to the barrier at the interface between the confined system and the rest of space. The GSWS potential has three parameters, $V_0$ measures the depth of the potential, the parameters $a$ and $L$ are used to manipulate the shape of the potential. As $a$ decreases, the surface contribution of the potential gets more extended and the bulk contribution gets sharper, while the parameter $L$ measures the nuclear size. The GSWS potential is used to simulate some physical problems, showing good agreement with the experiments [26-30].

In Figure 1, the GSWS potential for the nucleon confined within a Fluorine 17 nucleus is shown for the values of the parameters $V_0 = 42.710\ MeV$ and $a = 1.538\ fm^{-1}$[12], and for the calculated radius of the Fluorine 17 nucleus, $L = 3.209\ fm$. The GSWS potential is completely symmetric with respect to ordinate axis, thus, even and odd wave functions are expected for corresponding eigenstates.

The nucleon under consideration has the bound state energies satisfying $-V_0 < E_n^b < 0$, i.e. $-42.710\ MeV < E_n^b < 0$ and quasi-bound states satisfying $0 < E_n^{q\ b} < V_0 \frac{(1-aL)^2}{4aL}$, i.e. $0 < E_n^{q\ b} < 33.523\ MeV$. One dimensional Schrödinger equation for the nucleon with mass $m$ moving under the GSWS potential has been extensively studied by Lütfüoğlu et al. [14]. Employing the same procedure as in [14] for the parameters for Fluorine 17 nucleus, the energy spectrum is calculated as shown in Table. The energy spectrum together with the GSWS potential is demonstrated in Figure 1, and their corresponding unnormalized wave functions are plotted in Figure 2.

Note that the energy eigenvalues of bound states are real and they have infinite time constant, therefore the states are stationary. Contrarily, energy eigenvalues of quasi-bound states have complex form in general. This implies a finite time constant and thus a non-zero transition probability.

### 2.1. Thermodynamic functions of the system

In order to analyze the system in the context of thermodynamics, we first write the partition function by using the energy levels of the system

$$Z(\beta) = \sum_n e^{-\beta E_n}, \qquad (3)$$

where $\beta = \frac{1}{k_B T}$, $k_B$ is the Boltzman constant and $T$ is the temperature in Kelvin.

Helmholtz free energy $F(\beta)$ of a system is defined by

$$F(\beta) \equiv -\frac{1}{\beta} \ln Z(\beta). \qquad (4)$$

Entropy $S(T)$ of a system is given by the relation

$$S(T) = -\frac{\partial}{\partial T} F(T). \qquad (5)$$

The Helmholtz free energy and the entropy functions for both quasi-bound states included and excluded of the system vary with reduced temperature $k_B T/mc^2$, as seen in Figure 3 (a) and (b), respectively.

The zero entropy at zero Kelvin is in agreement with the third law of thermodynamics. Note that the entropy saturates to a higher value at a higher rate when quasi bound state is included in the partition function. This is attributed to the fact that inclusion of the bound states gives rise to an increase in the number of microstates available to the system. Since the Helmholtz free energy is equal to the minus integral of entropy with respect to temperature, when quasi bound state is included, its curve versus temperature goes below the one for bound states only.

The internal energy $U(\beta)$ of the nucleon is merely the quantum mechanical expectation value of the energy. In the language of statistical mechanics, the internal energy is related to the partition function by

$$U(\beta) = -\frac{\partial}{\partial \beta} \ln Z(\beta). \qquad (6)$$

The isochoric specific heat capacity $C_v(T)$ is defined by

$$C_v(T) \equiv \frac{\partial}{\partial T} U(T). \qquad (7)$$

The internal energy and the specific heat capacity functions for both quasi-bound states included and excluded of the system vary with temperature as seen in Figure 3 (c) and (d), respectively.

## 3. RESULTS AND DISCUSSION

At zero Kelvin, the starting value of internal energy is the lowest energy level of the spectrum $-28.297\ MeV$, as given in Table. The internal energy starts to increase at around 0.002 of the reduced temperature, where the probability of occupation the ground state energy lessens from unity. The internal energy keeps increasing in a convex manner until a certain reduced temperature about 0.013, at which the specific heat capacity passes through a maximum, followed by concave ascent up to the saturation as shown in Figure 4. At about huge reduced temperature of 1.53, the probability of occupation the first excited state passes through a maximum.

When the quasi-bound state is included in the partition function, the maximum in the specific heat capacity occurs at a higher value, at a higher temperature. This indicates that the internal energy reaches the saturation value in a longer temperature range than that only bound states are included.

Significant decay probability starts to arise at very high temperature around $10^{11}$K. As the temperature goes to infinity, all states become equally available and the internal energy goes to the arithmetic average value of the energy spectrum. The calculated quasi bound state energy $19.673 MeV$ is in reasonable agreement with the experimental data on the cumulative fission energy [32].

## 4. CONCLUSIONS

In this paper, we presented results of exact analytical solution of the Schrödinger equation for the GSWS potential for a nucleon in a Fluorine 17 nucleus. We calculated even and odd wave functions for bound and quasi-bound states, their corresponding energy eigenvalues, and the partition functions as functions of temperature. All calculations are carried out for the cases of quasi-bound state is included and excluded.

Using the partition functions in the two cases, we calculated the temperature dependent Helmholtz free energies, the internal energies, the entropies and the specific heat capacities of the system. When the quasi-bound state is included, the internal energy increases, the Helmholtz energy decreases, the entropy increases at high temperatures. The internal energy has an inflection point at the temperature about $1.4 \times 10^{11}$ K, at which the specific heat capacity passes through a maximum. Examining the wave functions, it is observed that the nucleon is completely restricted within the nucleus, i.e., no decay probability for bound states, while very low tunneling probability arises for quasi-bound state. This practically means that Fluorine 17 nucleus can only emits positron accompanied by x-ray, as confirmed by the experiments [32, 33]. This result is in agreement with the variation of the internal energy with temperature.

## ACKNOWLEDGEMENTS

This work was partially supported by the Turkish Science and Research Council (TUBITAK) and Akdeniz University.## REFERENCES

**Tables**

| $E_0^b$ | $E_1^b$ | $E_2^{q\,b}$ |
|---------|---------|--------------|
| -28.297 | -4.725  | 19.673-0.699 i |

**Table**. The calculated energy spectrum of Fluorine 17 nucleus in the unit of MeV. The highest level corresponds to quasi-bound eigenvalue. Note that the modulus of this energy is very close to its real part.

**Figures**

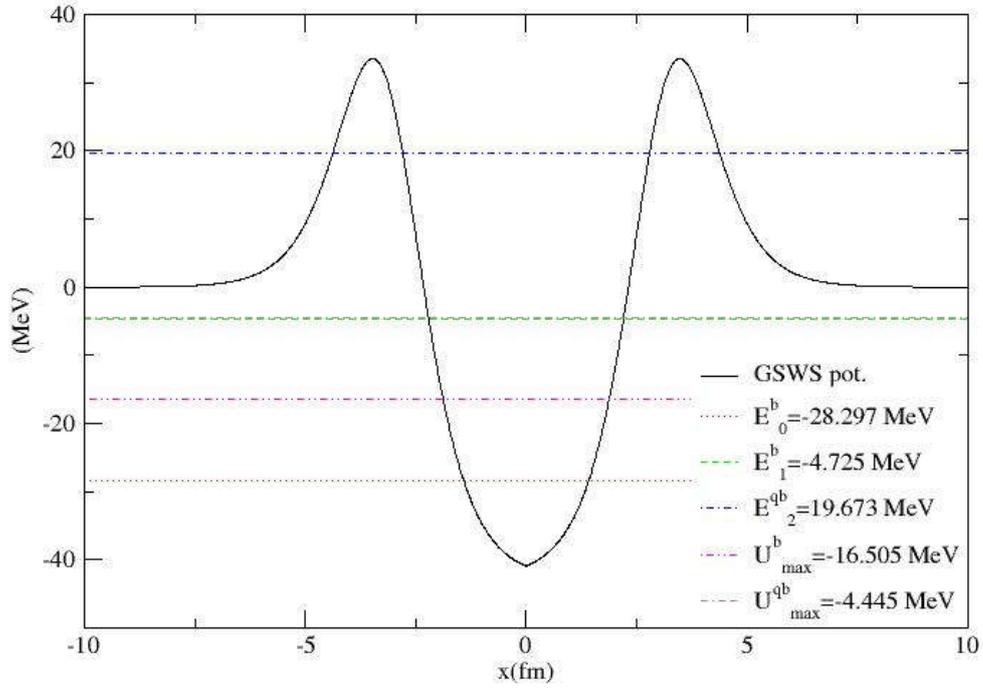

**Figure 1.** The GSWS potential and energy eigenvalues for the Fluorine 17 nucleus. The saturation values of the internal energy for quasi-bound states excluded ($U_{max}^{b}$) and included ($U_{max}^{q\,b}$).

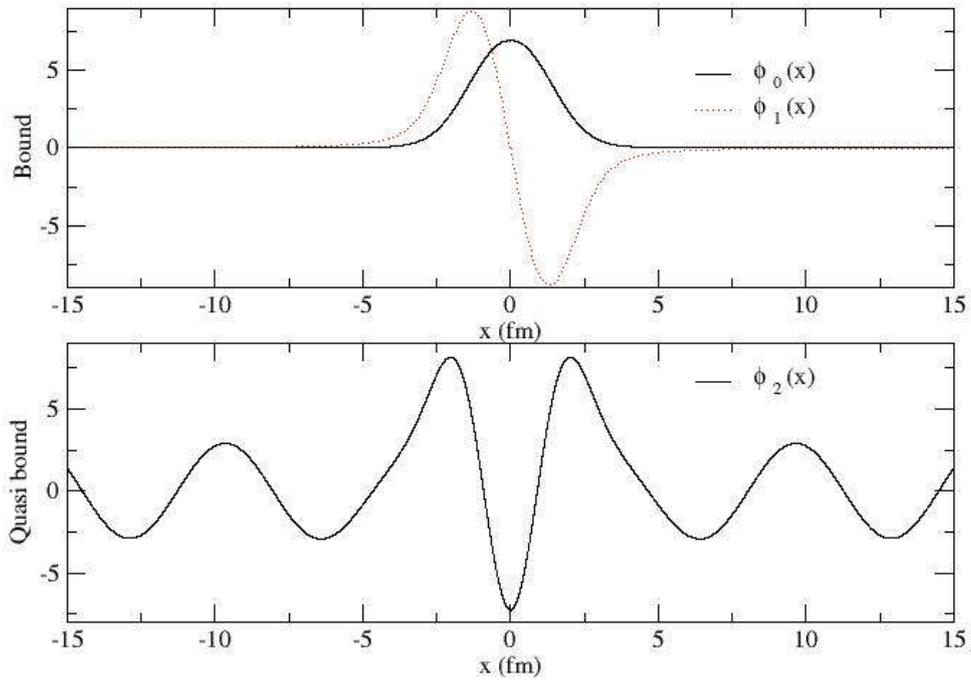

**Figure 2.** Even and odd unnormalized wave functions versus x for bound and quasi-bound states. Subscripts stand for the number of nodes. For quasi-bound state in the lower panel, tunneling indicated by the harmonic behavior outside the well is clearly observed. Note that one even and odd wave functions exist for bound states, while only one even for quasi bound state.

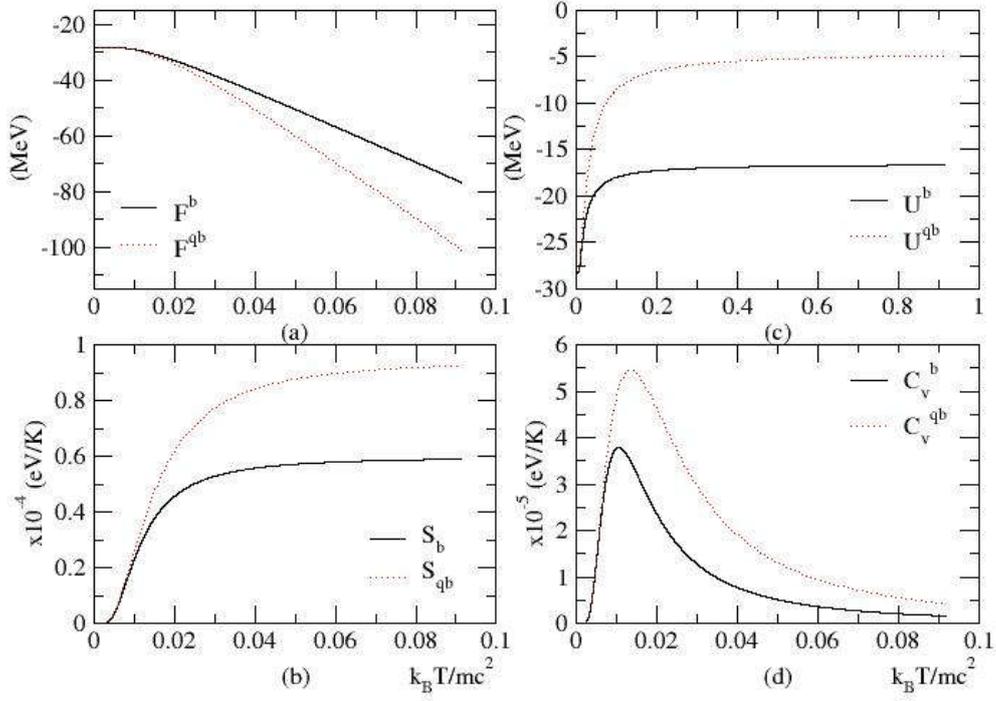

**Figure 3.** (a) Helmholtz free energy, (b) entropy, (c) internal energy, (d) specific heat capacity versus reduced temperature $k_B T/mc^2$. The quasi bound state is included in the red(dashed), excluded in the black(solid) curve.

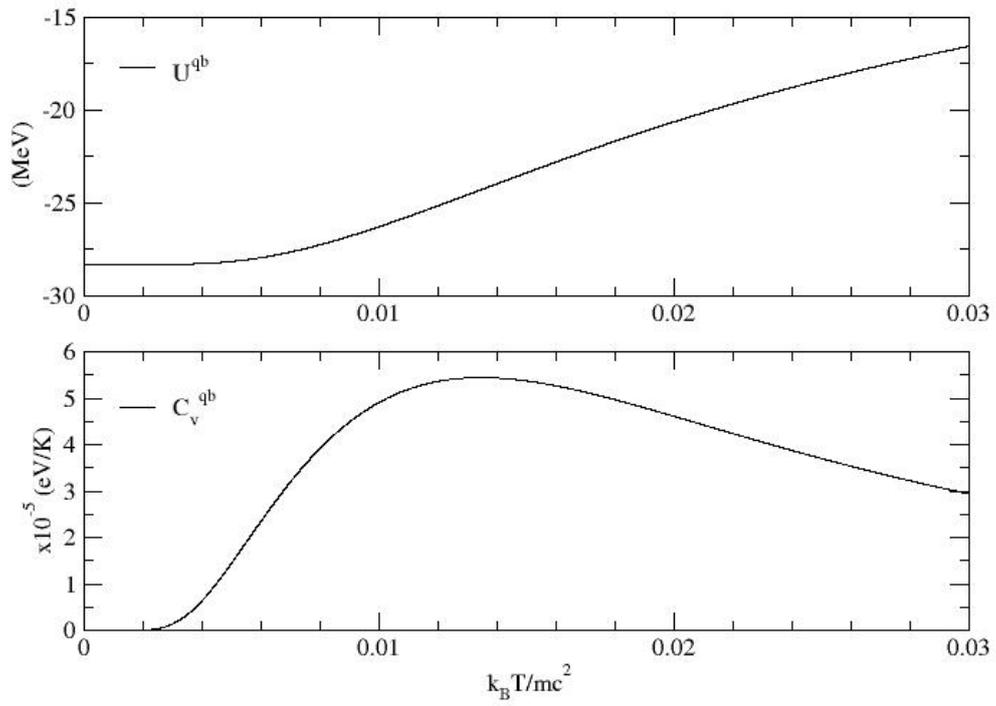

**Figure 4.** The internal energy has an inflection point at the first excited state, where the specific heat capacity passes through a maximum around zero temperature.